\documentclass{pasj00}
\draft

\begin{document}
\SetRunningHead{H. Tsunemi et al.}{Abundance Inhomogeneity in the South
East Limb of the Cygnus Loop} 
\Received{2008/08/14}
\Accepted{2008/10/09}

\title{Another Abundance Inhomogeneity in the South East Limb of
the Cygnus Loop}  

\author{Hiroshi \textsc{Tsunemi}\altaffilmark{1}, Masashi
  \textsc{Kimura}\altaffilmark{1}, Hiroyuki
  \textsc{Uchida}\altaffilmark{1}, Koji \textsc{Mori}\altaffilmark{2},\\
  and Satoru \textsc{Katsuda}\altaffilmark{3,1}}
\email{tsunemi@ess.sci.osaka-u.ac.jp, mkimura@ess.sci.osaka-u.ac.jp,
  uchida@ess.sci.osaka-u.ac.jp,\\ mori@astro.miyazaki-u.ac.jp, and
  Satoru.Katsuda@nasa.gov}
  %
\altaffiltext{1}{Department of Earth and Space Science, Graduate School of
  Science, Osaka University,\\
	 1-1 Machikaneyama, Toyonaka, Osaka
  560-0043, Japan}
\altaffiltext{2}{Department of Applied Physics, Faculty of Engineering, University of Miyazaki\\
1-1 Gakuen Kibana-dai Nishi, Miyazaki, 889-2192, Japan}
\altaffiltext{3}{Code 662, NASA Goddard Space Flight Center, Greenbelt, MD 20771,
U.S.A.}


%


\maketitle

\begin{abstract}
We have observed the south-east (SE) limb of the Cygnus Loop with {\it Suzaku}.  Our spatially-resolved spectroscopic study shows that a
one-$kT_\mathrm{e}$ non-equilibrium ionization model represents our spectra fairly well. We find that the metal abundances obtained are
all depleted relative to the solar values with a positional dependency
along the radial direction of the Cygnus Loop. The abundances in the
very edge of the limb shows about half the solar value, whereas
other regions inside the Loop show about 0.2\,times the solar value
which has been believed as a typical value for the Cygnus Loop limb. The
\lq\lq enhanced\rq\rq~abundance in the very edge in the SE limb is quite similar to that found
in the north-east (NE) limb of the Loop, and thus this is another evidence of
abundance inhomogeneity in the limb regions of the Loop.  The radio map shows a quite different feature: the NE limb is in the radio bright region while the SE limb shows almost no radio.  Therefore, the metal abundance variation in the SE limb can not attribute to the non-thermal emission.  The abundance inhomogeneity as well as the metal depletion down to 0.2\,times
the solar value still remain an open question.
\end{abstract}

\section{Introduction}

The Cygnus Loop is one of the brightest sources below 1\,keV.  It is a nearby (540\,pc: \cite{Blair2005}) proto-typical
middle-aged ($\sim$10,000 yr) supernova remnant (SNR) located near the galactic plane: ($l,b$)=(74$^\circ$, $-8^\circ.5$). 
The foreground neutral hydrogen column density, $N_\mathrm{H}$, is
estimated to be $\sim4\times10^{20} \mathrm{cm}^{-2}$
(\cite{Inoue1979}; \cite{Kahn1980}).  The low foreground absorbing
material as well as the large apparent size (2$^\circ$.5 
$\times$ 3$^\circ$.5: \cite{Levenson1997}; \cite{Aschenbach1999}) and high
surface brightness enable us to perform a detailed study of the soft X-ray emission from the
Cygnus Loop.   

The supernova (SN) explosion of the Cygnus Loop is generally considered
to have occurred in a preexisting cavity (firstly noted by
\cite{McCray1979}).  The X-ray boundary in the northeastern (NE) limb of the Cygnus Loop is associated with Balmer-dominated filaments which mark current locations of the blast wave (\cite{Chevalier1978}).  Hester et al.\ (1994) studied a Balmer-dominated filament in detail and reported that it was recently (in past 1000 years) decelerated from $\sim$400 to $\sim$180\,km\,s$^{-1}$.  This rapid deceleration of the shock velocity was considered to be a result of the blast wave hitting the wall of a cavity surrounding the supernova precursor.  
Levenson et al.\ (1999) observed the whole remnant
by {\it ROSAT} PSPC and generated the softness map.  They found a soft spatially thin ($<5^\prime$) shell around almost the entire limb of the Loop. 
Based on the observation, they concluded that the soft shell occurred where the cavity walls of the Cygnus Loop decelerated the blast wave.


{\it ASCA} observation of the NE limb of the Cygnus Loop revealed the non-equilibrium ionization (NEI) plasma condition as well as the depleted metal abundances relative to the solar values by a factor of $\sim$5 (\cite{Miyata1994}, \cite{Miyata1998}).  The low metal abundances lead them to consider that the plasma in the NE-limb of the Cygnus Loop originated from the swept-up interstellar
medium (ISM) rather than the ejecta-material.  Recently, Miyata et al.\ (2007) observed the same region by {\it Suzaku}
(\cite{Mitsuda2007}) Science Working Group (SWG) time that enabled us to observe emission lines from highly ionized C and N for the first time and obtained depleted abundances: C$\sim$0.1, N$\sim$0.03, O$\sim$0.1, Ne$\sim$0.1, Mg$\sim$0.2, Si$\sim$0.4, Fe$\sim$0.2.  They confirmed the metal deficiency as well as the NEI condition there.   
Leahy (2004) observed the southwestern (SW) limb of the Cygnus Loop by the {\it Chandra} satellite and performed spatially resolved spectral analysis.  In his spectral analysis, he divided the elements into three groups: O-group, Ne-group and Fe-group.  He found the collisional ionization equilibrium (CIE) condition and deficient metal abundances (mean O abundance is 0.22).  Levenson et al. (2005) observed the southeast (SE) region where a complicated intensity structure is seen.  They found the metal abundance to be 0.3--0.6 with collisional ionization equilibrium condition.  Therefore, the low metal abundances may be a common feature except the south east region of the Cygnus Loop.

Cartledge et al.\ (2004) measured the interstellar oxygen along 36 sight
lines and confirmed the homogeneity of the O/H ratio within 800\,pc of
the Sun.  We found that the closest sight line to the Cygnus Loop was
about $5^\circ$ away.  The oxygen abundance they measured is about 0.4
times the solar value (\cite{Anders1989}).  Wilms et al.\ (2000)
employed 0.6 of the total interstellar abundances for the gas-phase ISM
oxygen abundance, and suggest that this depletion may be due to grains.
Although the ISM near the Cygnus Loop may be depleted by a factor of 2
compared with the solar values, the abundances are still much higher
than that reported at the limb of the Cygnus Loop. It is difficult to
explain such a low abundance of oxygen in material originating from the
ISM. Therefore, the origin of the low metal abundance is not yet
answered. 

Recently, Katsuda et al.\ (2008a) found that the northern outermost
region in the NE limb of the Cygnus Loop showed enhanced abundances:
C$\sim$0.7, N$\sim$0.7, O$\sim$0.4, Ne$\sim$0.6, Mg$\sim$0.3, and
Fe$\sim$0.3 times the solar values.  The abundance enhanced region is about 30$^\prime \times 3^\prime$ region with relatively weak surface
brightness.  This is the outermost region of the Loop.  The inner part
of it showed depleted abundance.  Katsuda et al. (2008b) also confirmed
it by {\it Chandra}.  Based on the study of the plasma structure, they
concluded that neither a circumstellar medium, fragments of ejecta, nor
abundance inhomogeneities of the local interstellar medium around the
Cygnus Loop can explain the relatively enhanced abundance in the region.
The apparent abundance in the metal enhanced region of the NE limb of
the Cygnus Loop can be understood as the average metal abundance of the
ISM there.  Then they studied the possibility that the low metal
abundance may be due to the contribution of the non-thermal emission since the NE limb shows radio bright region (\cite{Uyaniker2004}).  The Cygnus Loop is one of the radio bright sources while it is not uniform.  The SE limb region of the Cygnus Loop shows no radio intensity.

We report here the
result of the observation of the SE limb of the Cygnus Loop by {\it Suzaku} showing a similar structure in abundance to that in the NE limb.

\section{Observations and Data Screening}

\begin{figure}
  \begin{center}
\includegraphics*[width=8cm, bb=0 0 568 649]{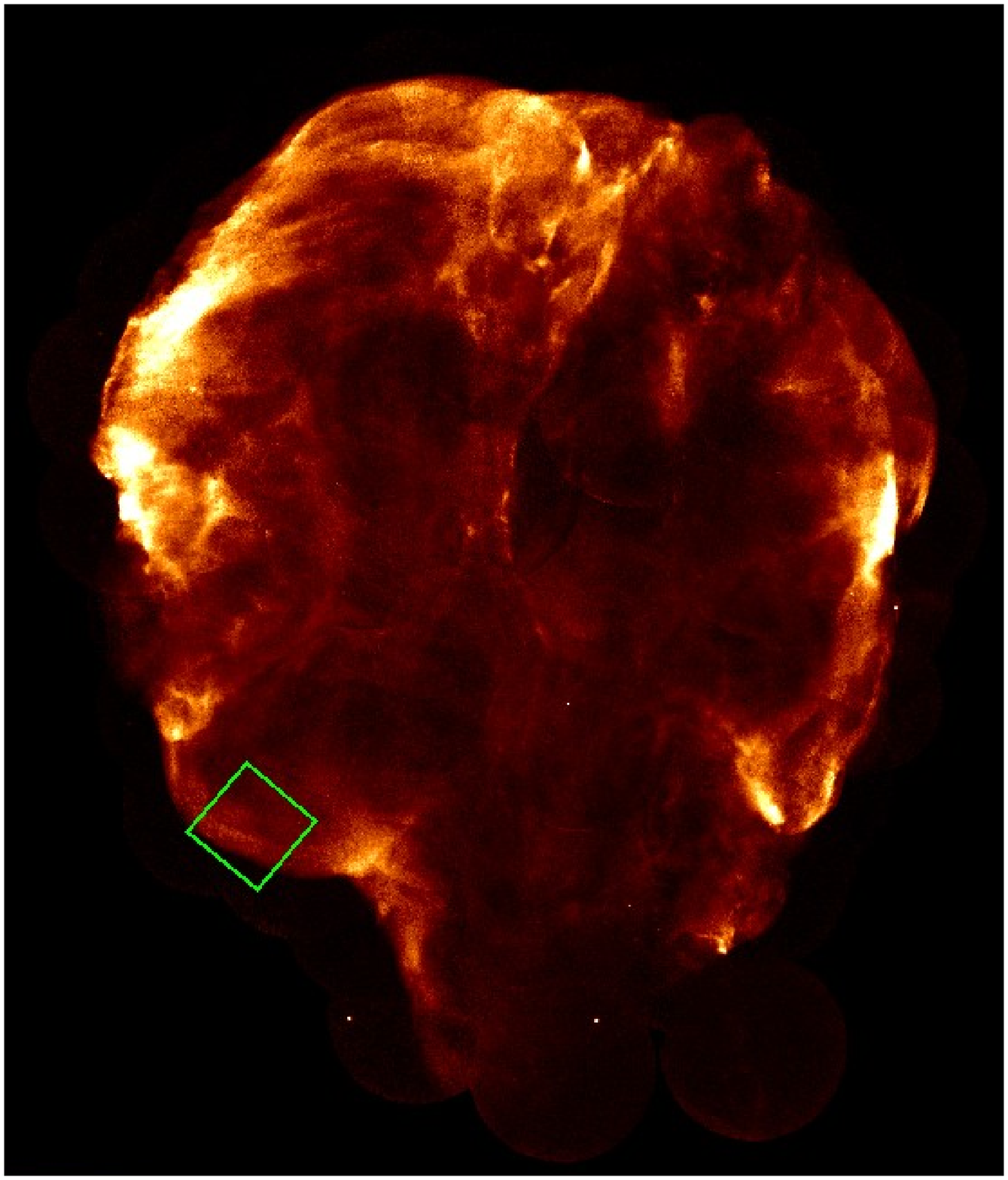}
\end{center}
  \caption{{\it ROSAT} HRI image of the entire Cygnus Loop. The {\it Suzaku} FOV in the SE limb is shown in green square. }\label{fig:hri_image}
\end{figure}

The observation was performed on 2008 May 12.
Fig.~\ref{fig:hri_image} shows the location of it by a square superposed
on the {\it ROSAT} image.  We employed revision 2.2 of the cleaned event data
and excluded the time region where the attitude was unstable.
Furthermore, we excluded data taken in the low cut-off rigidity
$<$\,6\,GeV\,c$^{-1}$.  The net exposure time was 18.9\,ks after the
screening.  Since the Cygnus Loop almost extends the entire FOV, we
subtracted a blank-sky spectrum obtained from the Lockman Hole whose observation date was close to that of
the Cygnus Loop (both are done in 2008 May).  We found no emission above 3\,keV within the
statistical uncertainties.

\begin{figure}
  \begin{center}
\includegraphics*[width=6cm, bb=0 0 609 599]{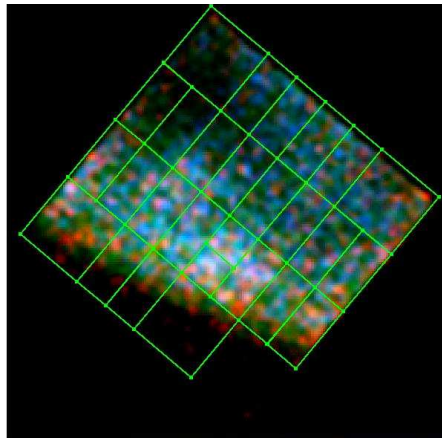}
  \end{center}
  \caption{Merged Suzaku XIS1 three color image of the SE limb (Red: 0.31--0.38\,keV band, Green: 0.38--0.46\,keV band,  Blue: 0.60--0.69\,keV band).  The effect contamination is corrected.	 We divided our FOV into 26 rectangles as indicated.}
	\label{fig:xis_image} 
\end{figure}

The XIS has a contamination problem (\cite{Koyama2007}) that a contaminant is accumulated on the optical blocking filter just in front of the CCD.  Although the increase of the contaminant almost terminates, the low energy response reduces more at our observation time than that at the SWG time.  The {\it Suzaku} observation of the NE limb of the Cygnus Loop was done for the first time in November, 2005 while our observation on the SE limb was carried out in May, 2008.  Due to the reduction of the response at low energy, we used photons in the energy range of 0.3--3.0\,keV and 0.5--3.0\,keV for XIS1 (back-illuminated CCD; BI CCD) and XIS0, 3 (front-illuminated CCD; FI CCD), respectively.  Figure~\ref{fig:xis_image} shows the {\it Suzaku} XIS1 three color image.  Red, green and blue represent the energy range of 0.31--0.38\,keV band, i.e., C {\scshape VI} K$\alpha$, 0.38--0.46\,keV band, i.e., N {\scshape VI} K$\alpha$, 0.60--0.69\,keV band, i.e., O {\scshape VIII} K$\alpha$, respectively.  In the data analysis, we grouped the spectra into bins of a minimum of 20 counts so that the errors are normally distributed to perform a $\chi^2$~test.  Therefore, the degree of freedom (d.o.f.) depends on the statistics of the data.

\section{Spatially Resolved Spectral Analysis}

We divided the entire FOV into 28 square/rectangular cells shown in Fig.~\ref{fig:xis_image} such that each cell has similar statistics.  We excluded the lower right square cell since it is almost out of the Loop.  We extracted spectra from the cells and performed spectral analysis.  In order to generate the response matrix file (RMF) and the ancillary response file (ARF), we employed {\tt xisrmfgen} and {\tt xissimarfgen} (version March 2008), respectively.  The low energy efficiency of the XIS shows degradation caused by the contaminants, however it was taken into account in the generation of the ARF file.  The latest version of the response includes all the available information after the launch (cf. Katsuda et al., 2008a).

\begin{figure}
  \begin{center}
\includegraphics*[width=12cm, bb=0 0 693 507]{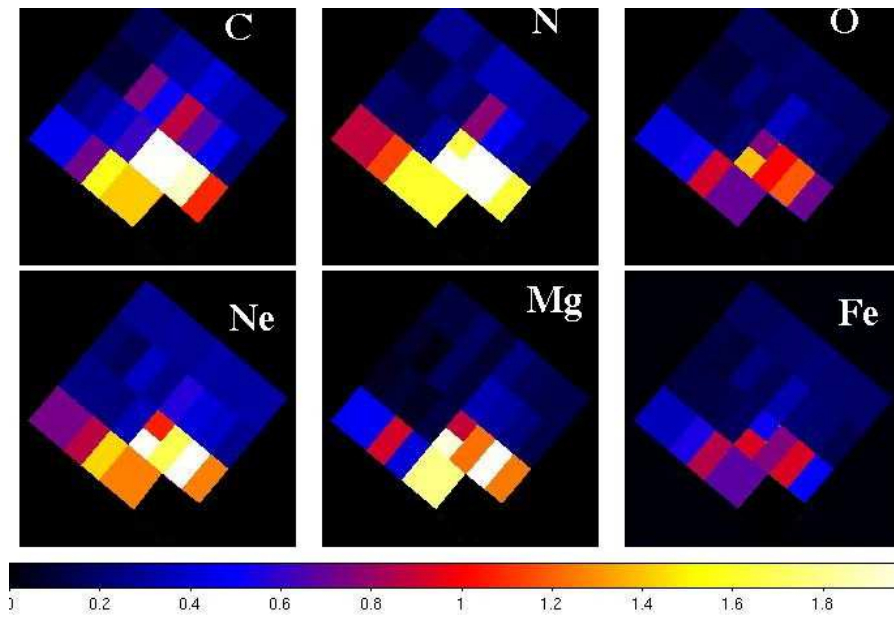}
  \end{center}
  \caption{Best-fit parameter maps are shown for various metal abundances.}
	\label{fig:map_param} 
\end{figure}

First of all, we applied an absorbed one-$kT_\mathrm{e}$ component of non-equilibrium ionization (NEI) model for all the spectra (the Wabs and VNEI model in XSPEC v\,12.4.0; \cite{Morrison1983}, \cite{Hamilton1983}, \cite{Borkowski1994}, \cite{Borkowski2001}, \cite{Liedahl1995}).  We noticed that many emission line features are seen while that emissions around 2\,keV or higher are very weak.  Therefore, free parameters employed are interstellar absorption feature, $N_\mathrm{H}$; electron temperature, $kT_\mathrm{e}$; the ionization time, $\tau$; the emission measure, EM (EM$=\int
n_\mathrm{e}n_\mathrm{H} dl$, where $n_\mathrm{e}$ and $n_\mathrm{H}$
are the number densities of electrons and protons, respectively and
$dl$ is the plasma depth); abundances of C, N, O, Ne, Mg, Si, S, Fe,
and Ni.  We set the abundance of Ni equal to that of Fe and those of
Si/S equal to that of O.  This
model gave us fairly good fits for all the spectra (reduced $\chi^2 <
1.3$).  Maps of the best-fit parameter values are presented in
Fig.~\ref{fig:map_param}.

Our data clearly show the NEI condition rather than the collisional ionization equilibrium (CIE) condition.  If we set log\,$\tau$= 12 where the CIE condition is achieved, we found that the reduced $\chi^2$ increased by $\sim$0.4.  Furthermore, we tried to fit data by using a CIE model (VMEKAL in XSPEC) which also shows the increase of the reduced $\chi^2$ by $\sim$0.3 than that of the best fit NEI model.  Therefore, the spectrum in our FOV surely shows the NEI condition.  

\begin{figure}
 \begin{center}
\includegraphics*[width=8cm, bb=0 0 570 563]{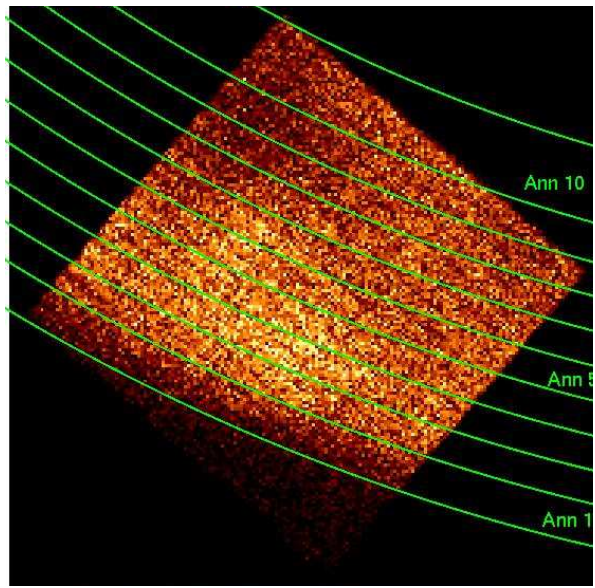}
  \end{center}
  \caption{10 annuli regions are shown which is employed for detailed spectral analysis.}
	\label{fig:ann_division} 
\end{figure}

\begin{figure}
  \begin{center}
\includegraphics*[width=8cm, bb=0 0 850 680]{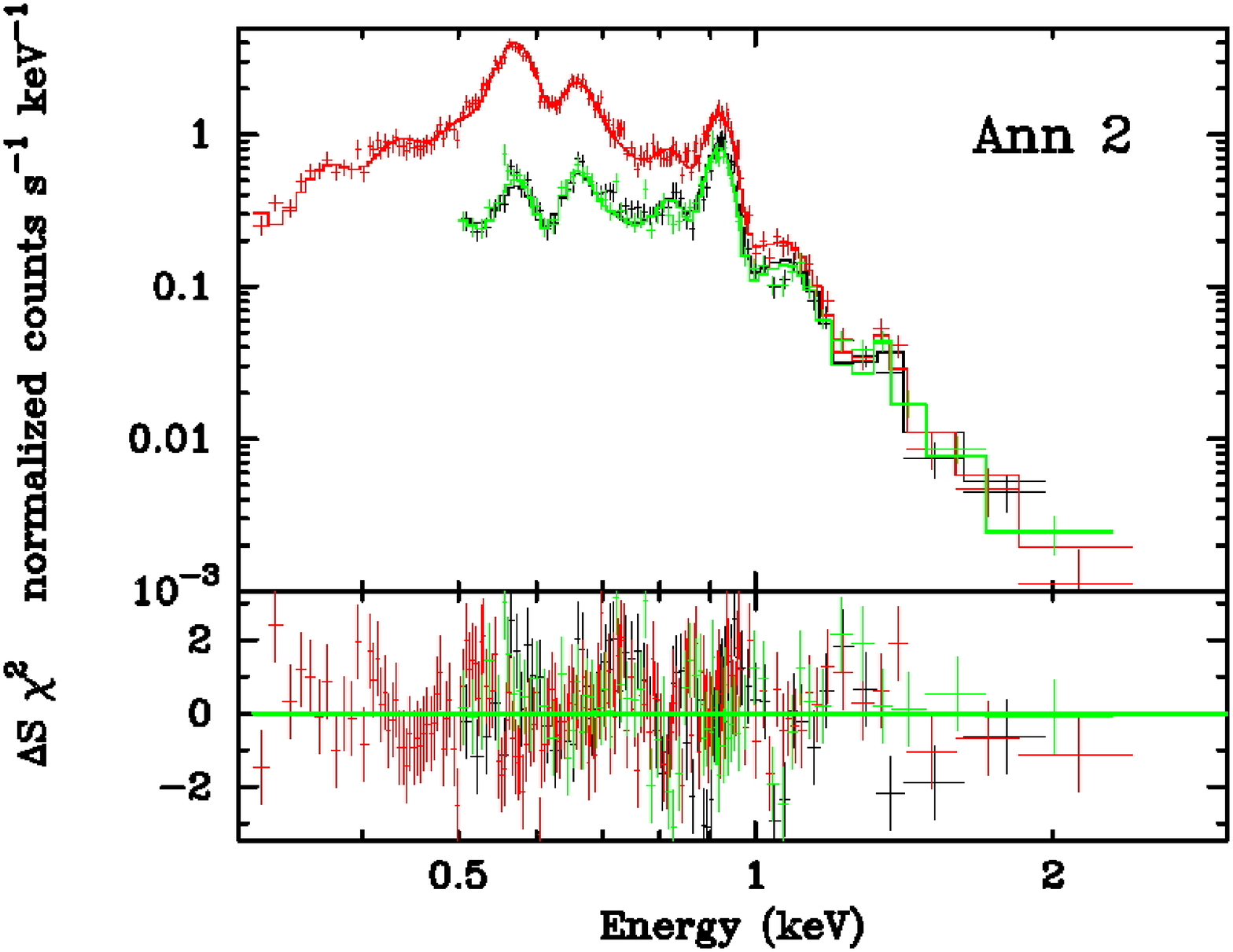}
\includegraphics*[width=8cm, bb=0 0 850 680]{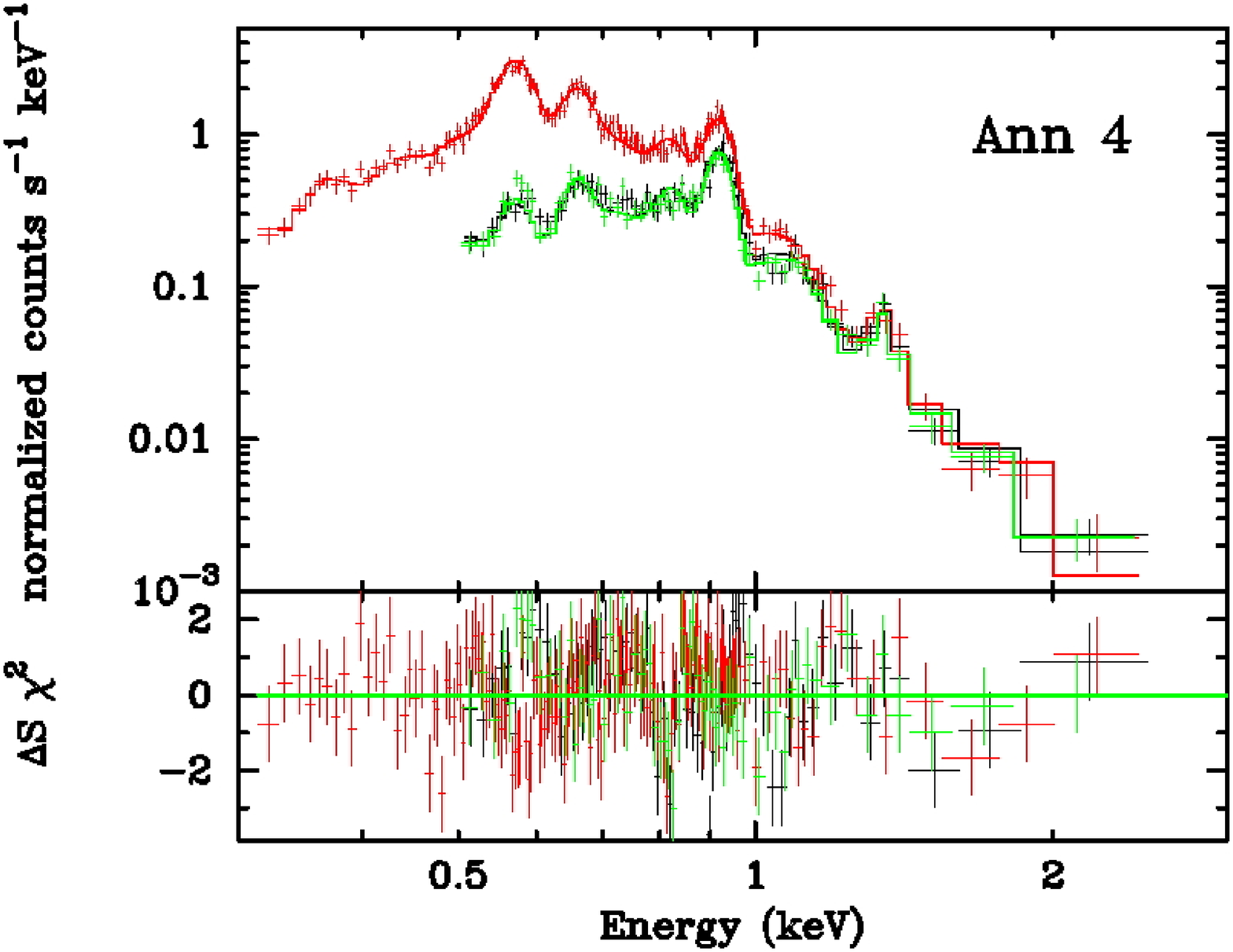}
\includegraphics*[width=8cm, bb=0 0 850 680]{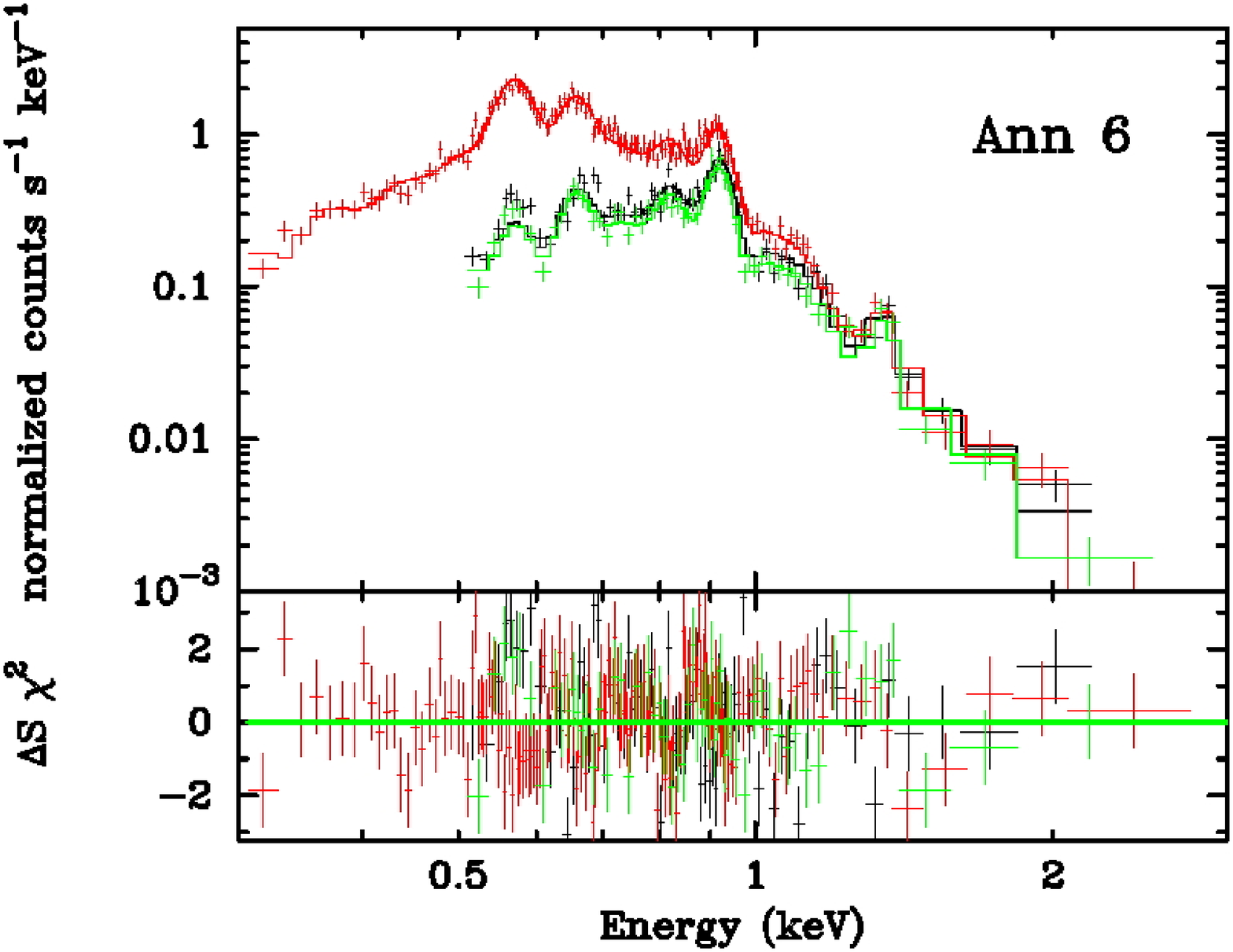}
\includegraphics*[width=8cm, bb=0 0 850 680]{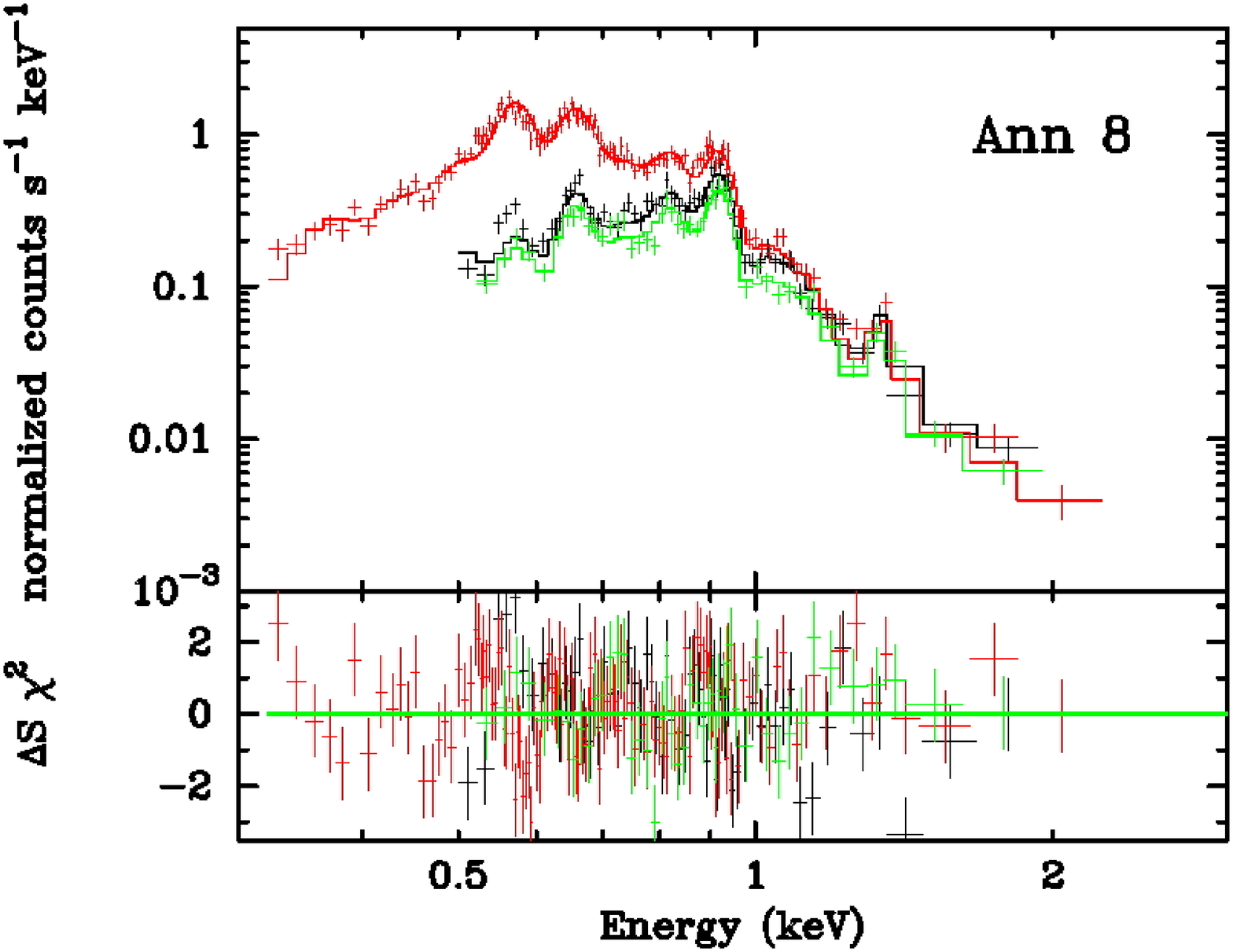}
  \end{center}
  \caption{Example spectra obtained in Ann02, Ann04, Ann06, Ann08 are shown with their best fit models.	}
	\label{fig:ex_spec} 
\end{figure}

We noticed that the limb regions show higher abundance while the temperature almost stays constant.  Furthermore, they show radial structure rather than the azimuthal structure.  Therefore, we have re-divided the FOV into 10 annular regions from Ann1 to Ann10 where the Ann1 is the outer most region as shown in Fig.~\ref{fig:ann_division}.  In this way, we performed the detailed spectral analysis employing the same model described before.  Fig.~\ref{fig:ex_spec} shows example spectra and their best fit models.  The reduced $\chi^2$ values are 1.2--1.7 with 250--440 d.o.f.  The detailed parameters are summarized in Table~\ref{tab:ex_param}.

\begin{figure}
  \begin{center}
\includegraphics*[width=5cm, bb=0 0 850 680]{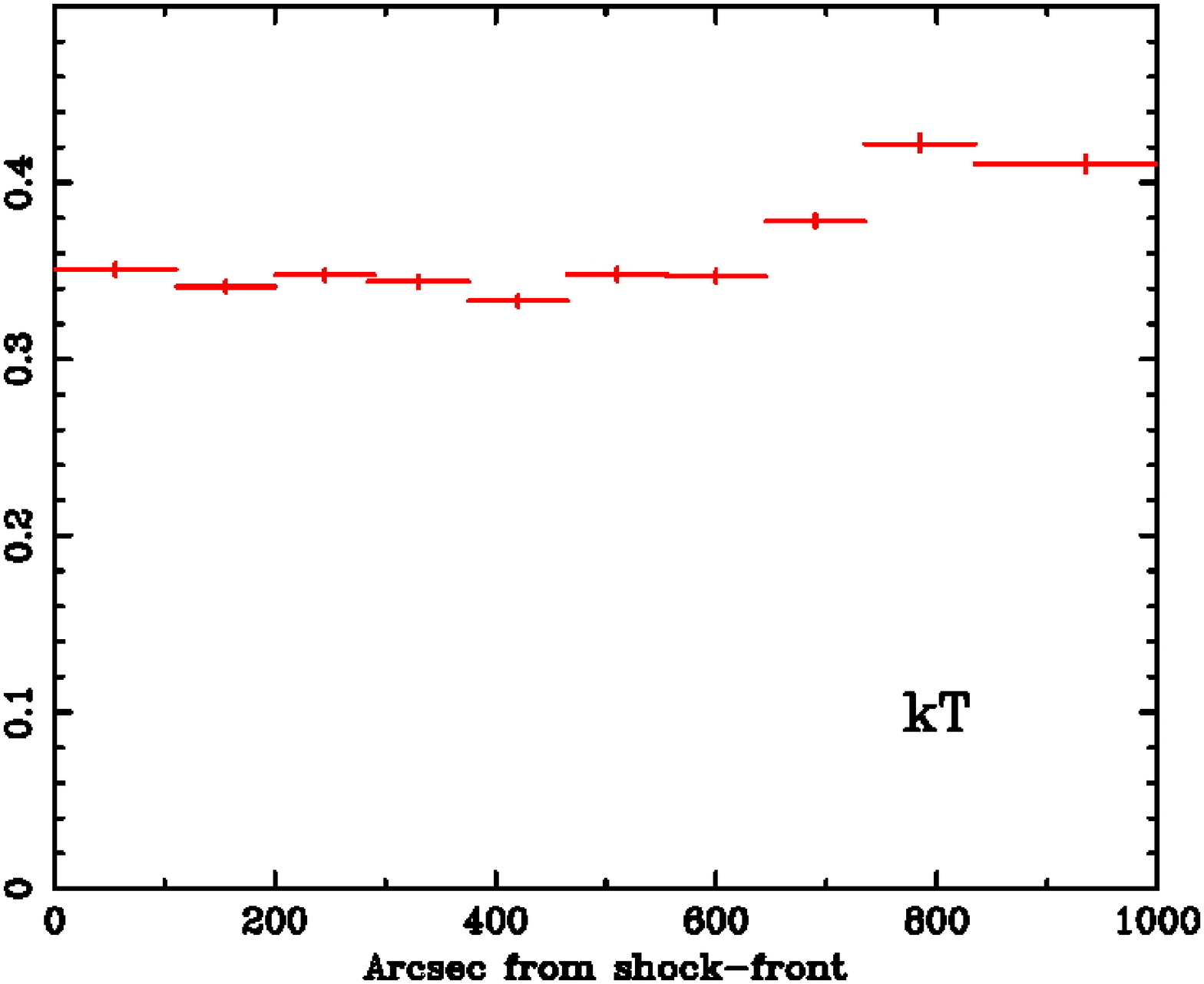}
\includegraphics*[width=5cm, bb=0 0 850 680]{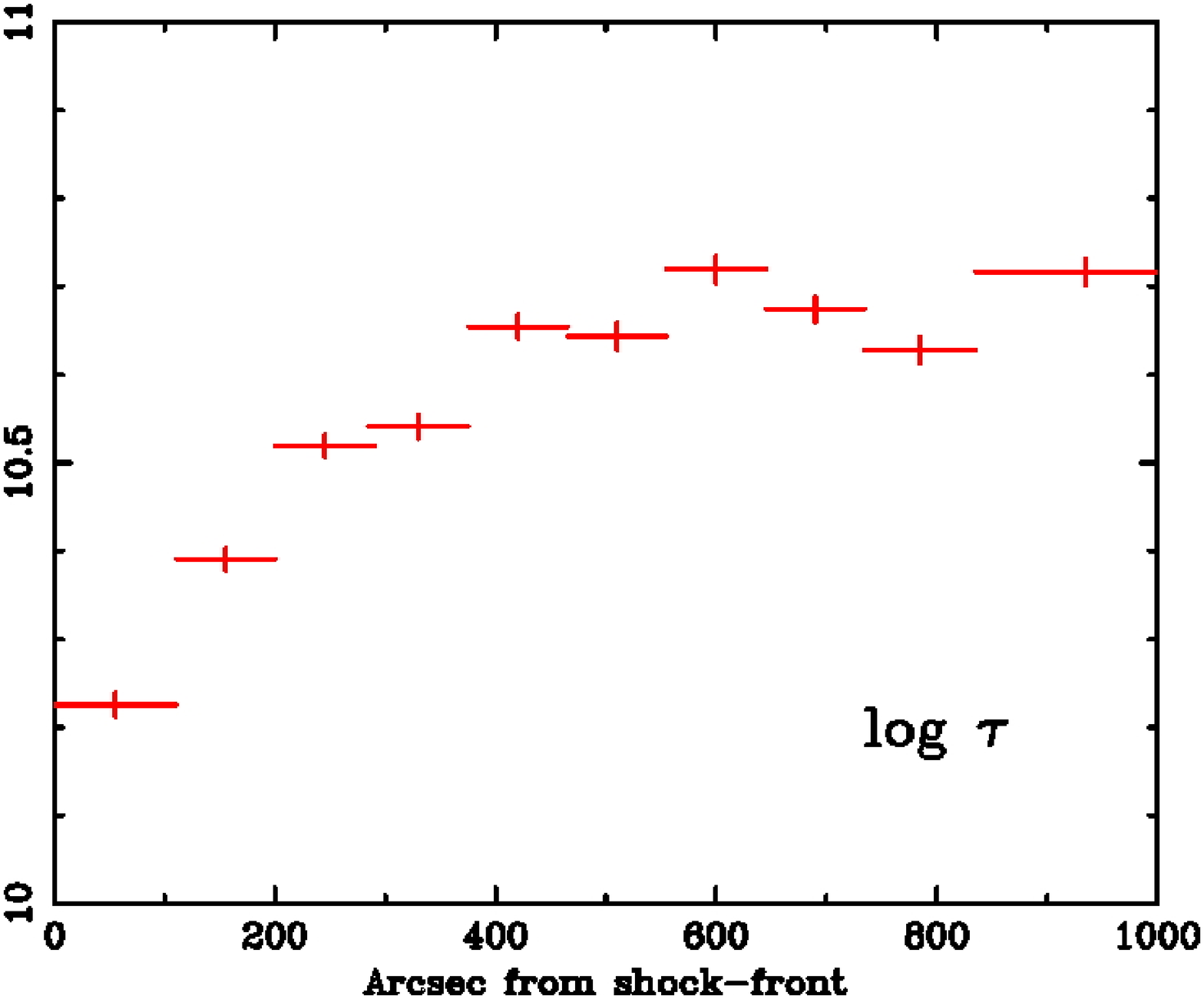}
\includegraphics*[width=5cm, bb=0 0 850 680]{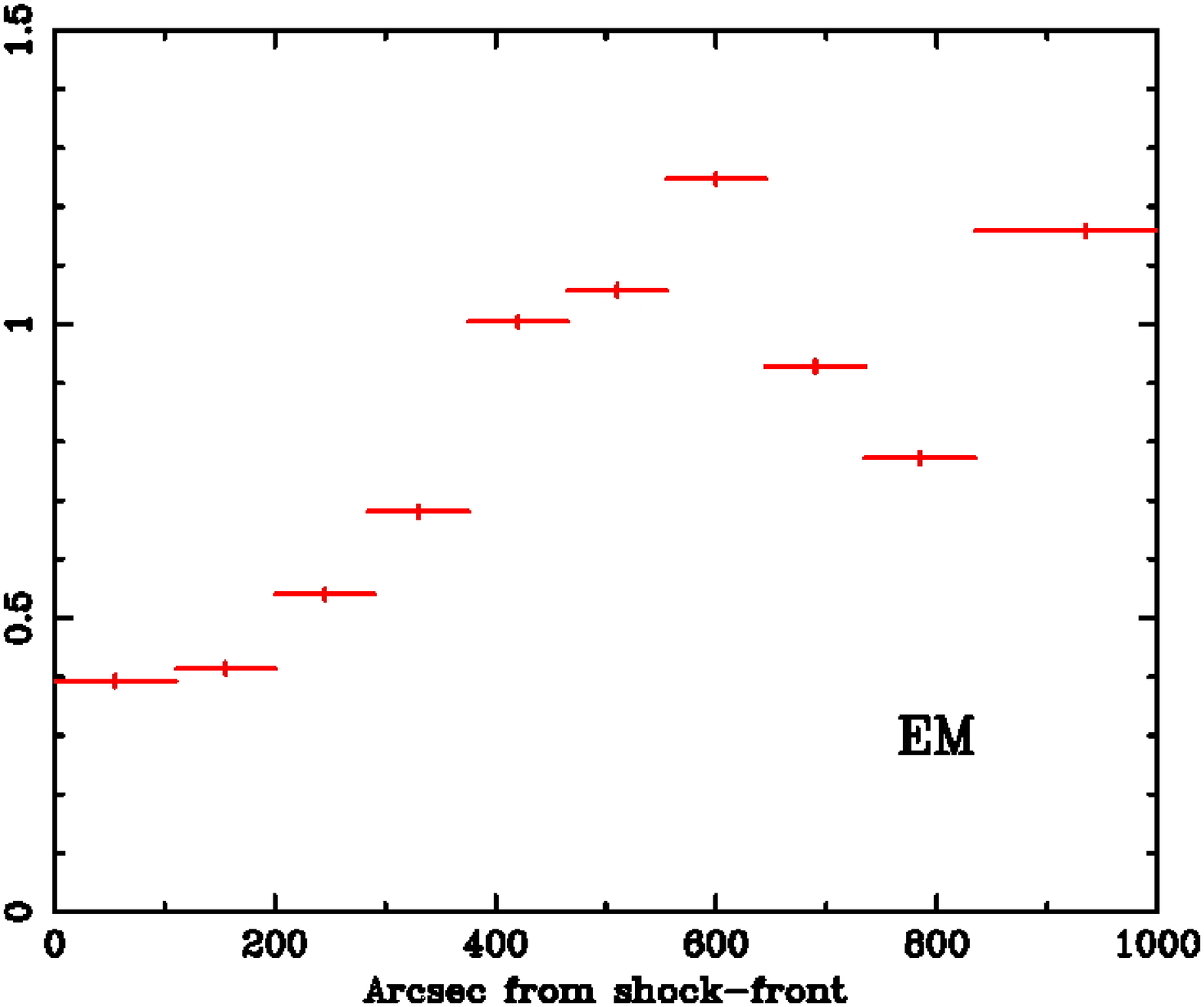}
\includegraphics*[width=5cm, bb=0 0 850 680]{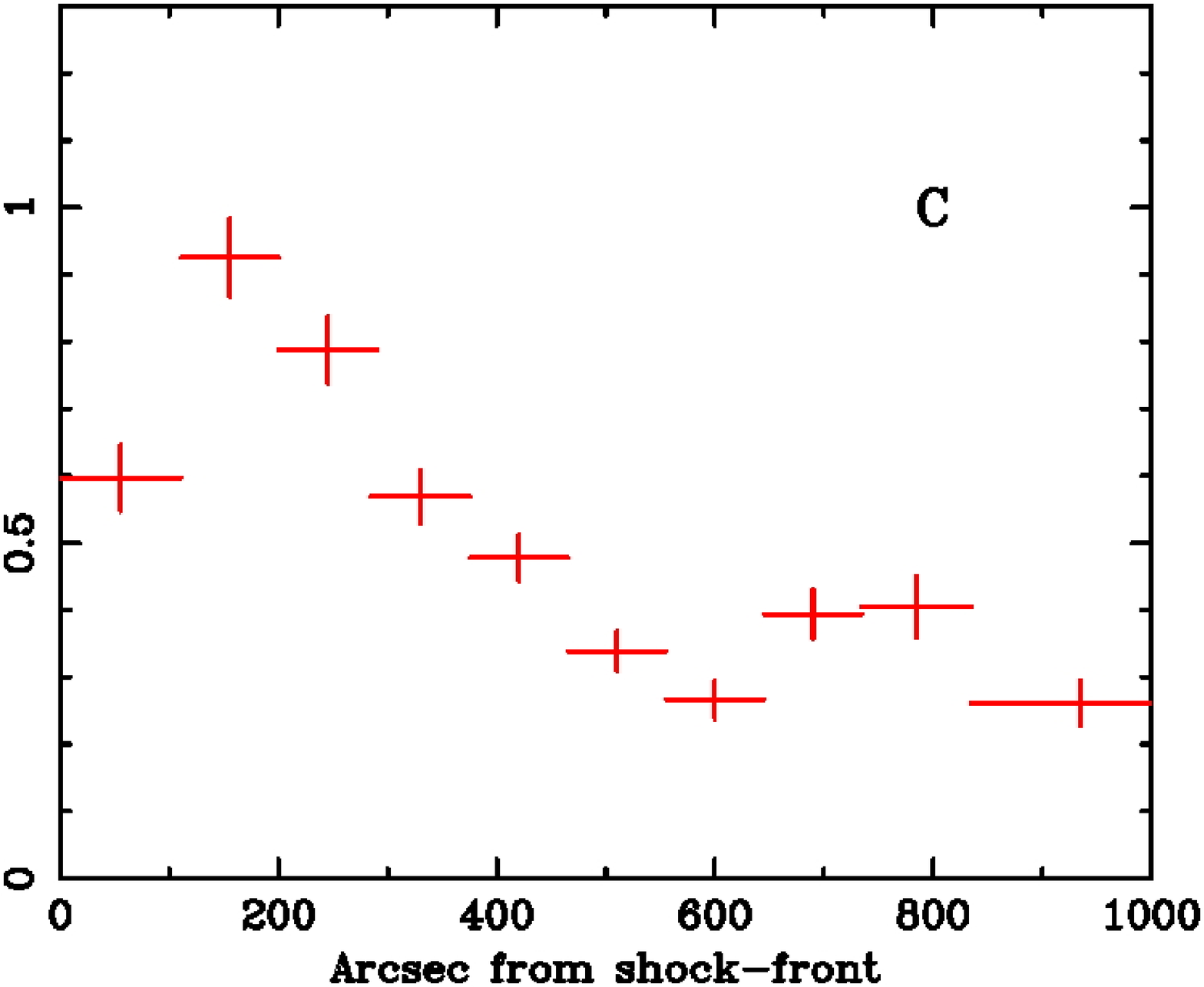}
\includegraphics*[width=5cm, bb=0 0 850 680]{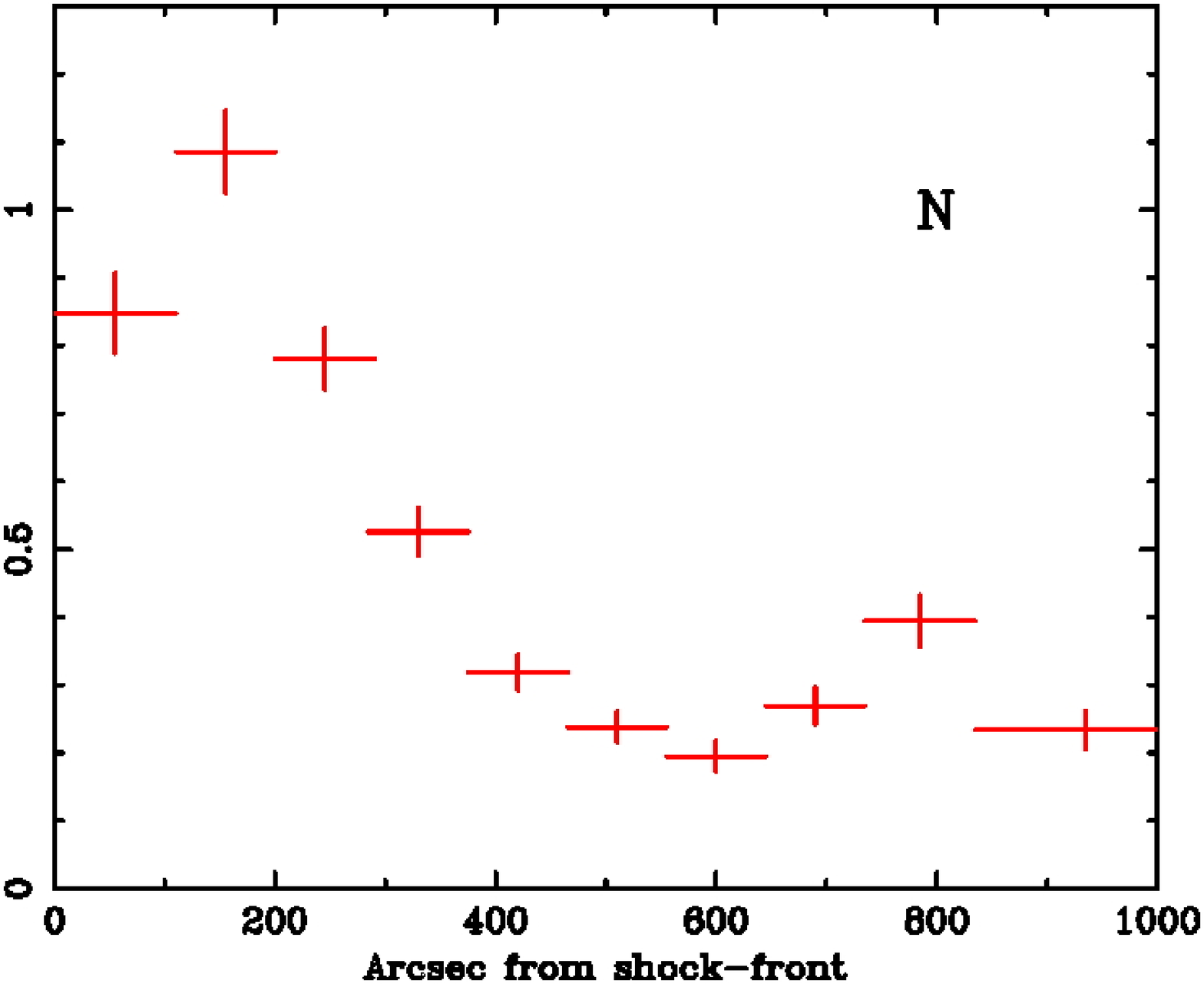}
\includegraphics*[width=5cm, bb=0 0 850 680]{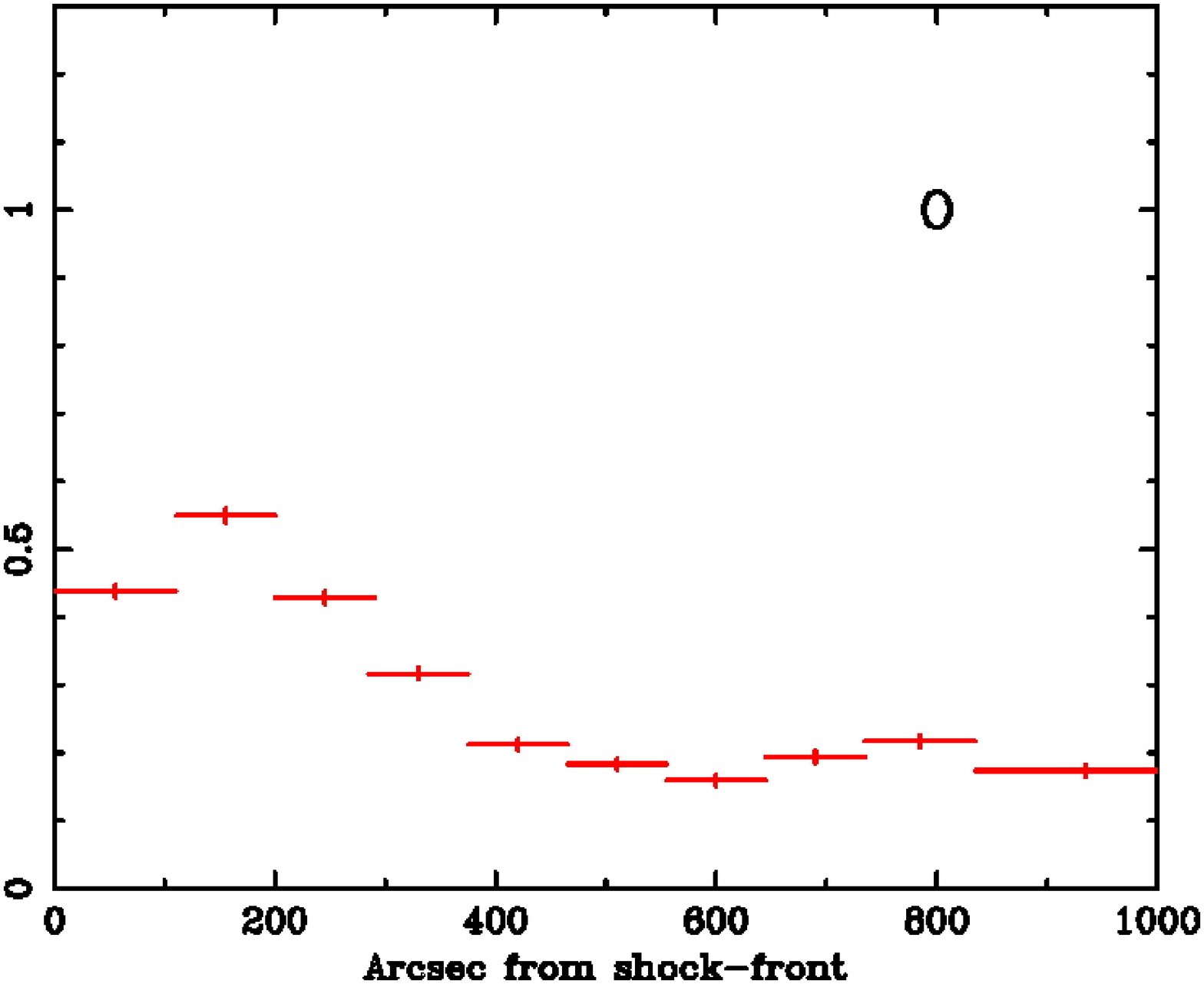}
\includegraphics*[width=5cm, bb=0 0 850 680]{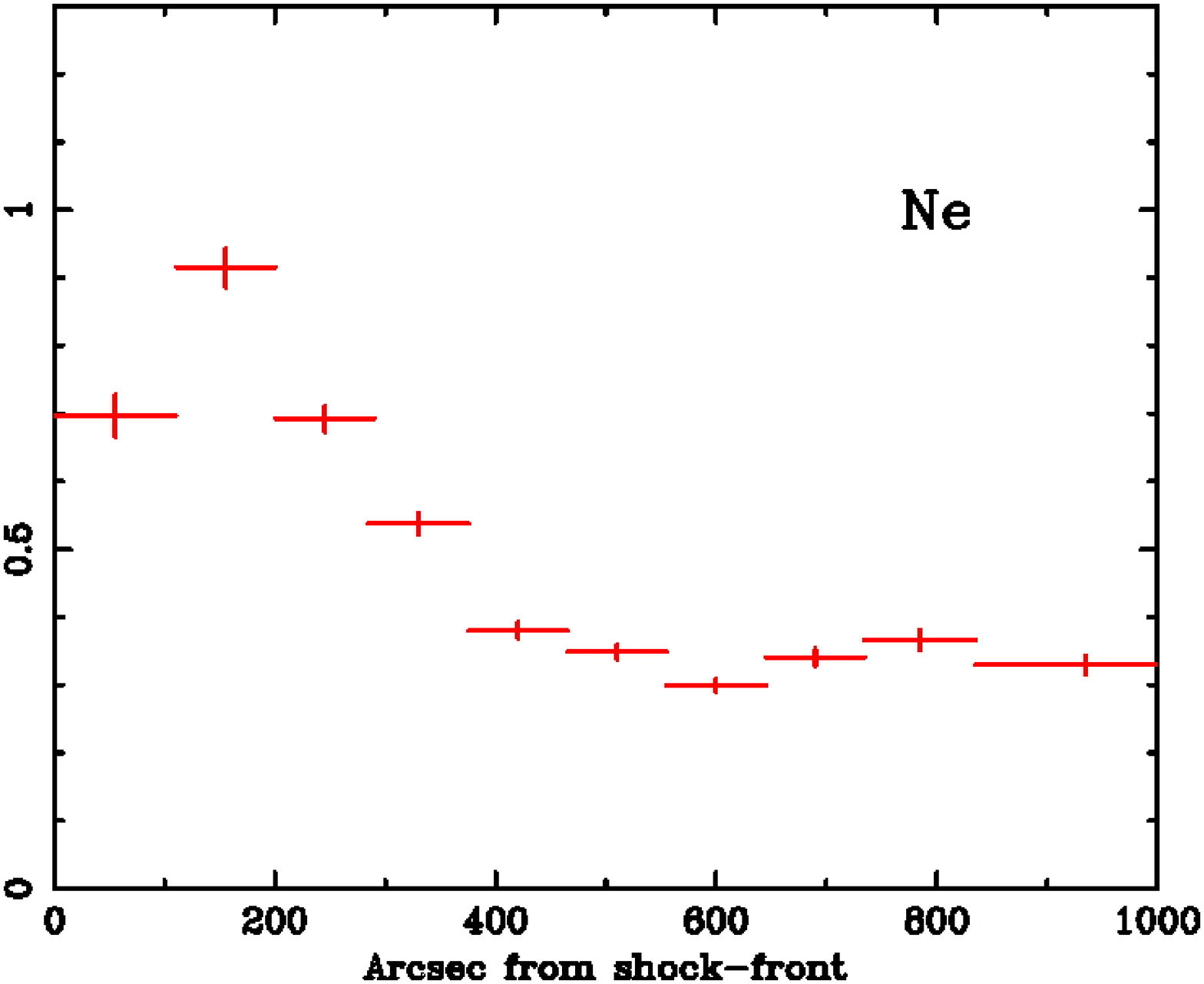}
\includegraphics*[width=5cm, bb=0 0 850 680]{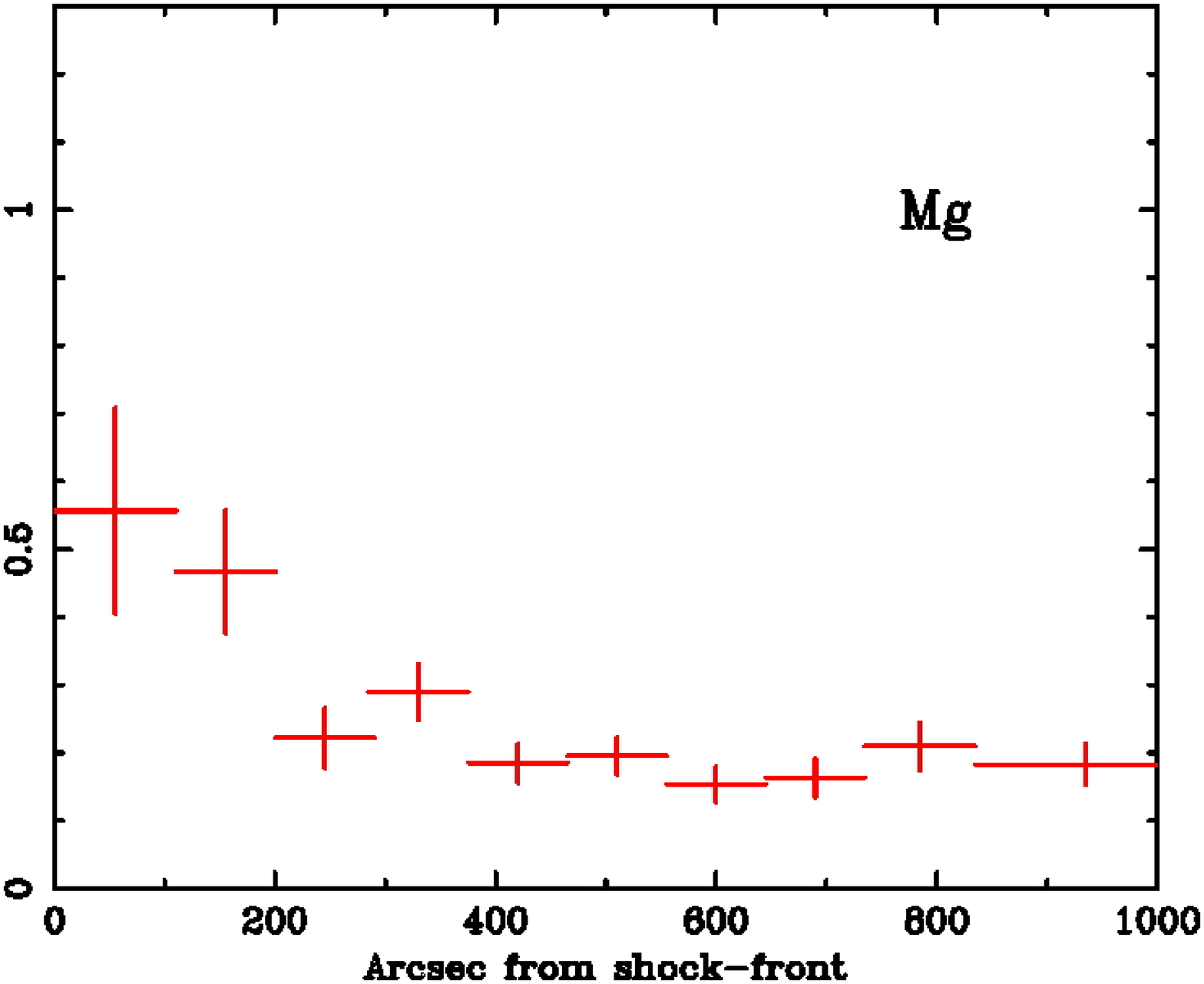}
\includegraphics*[width=5cm, bb=0 0 850 680]{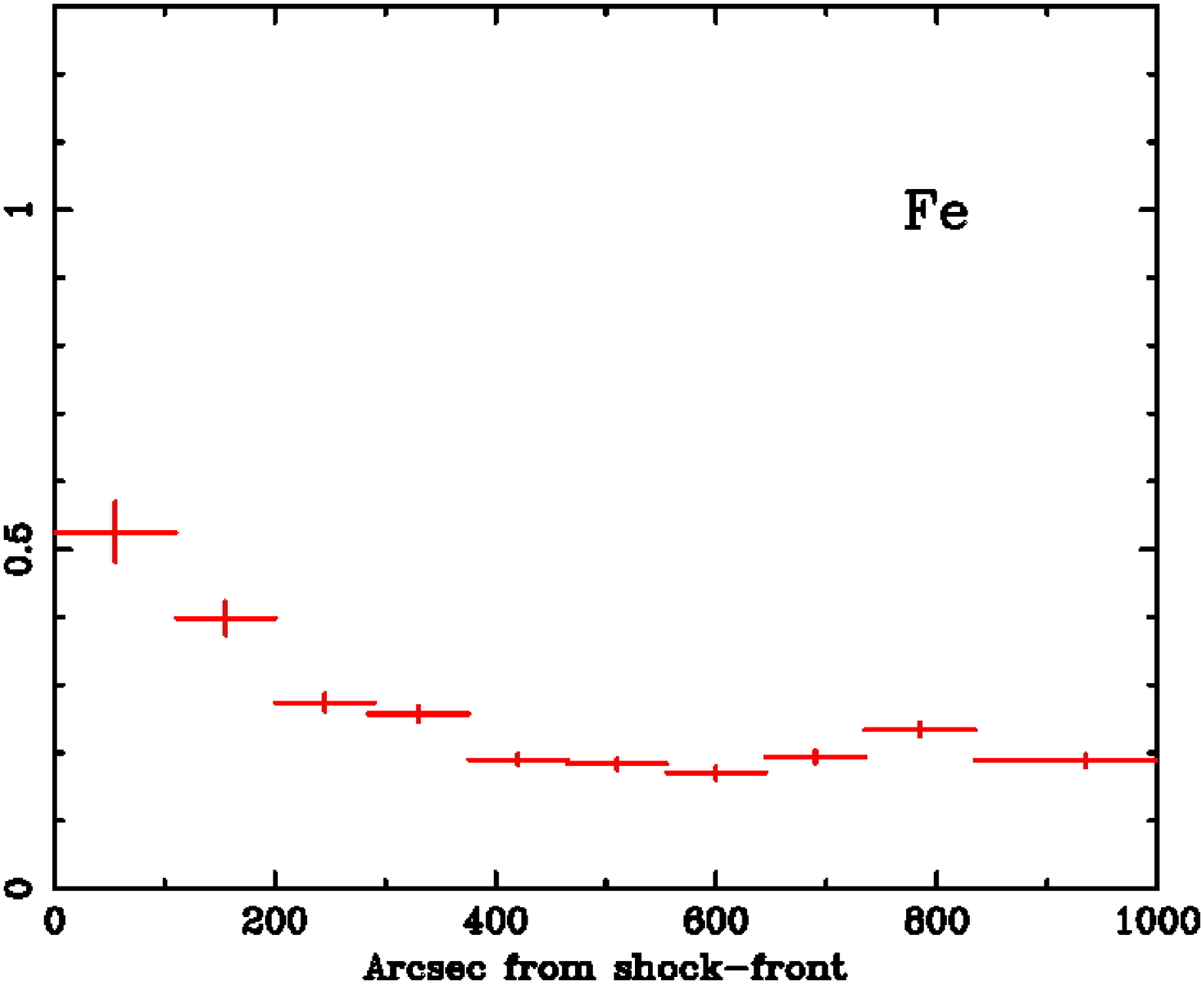}
  \end{center}
  \caption{Variations of various parameters are shown as a funciton of angular distance from the shock front.}
	\label{fig:correlation} 
\end{figure}

\begin{figure}
  \begin{center}
\includegraphics*[width=8cm, bb=0 0 850 680]{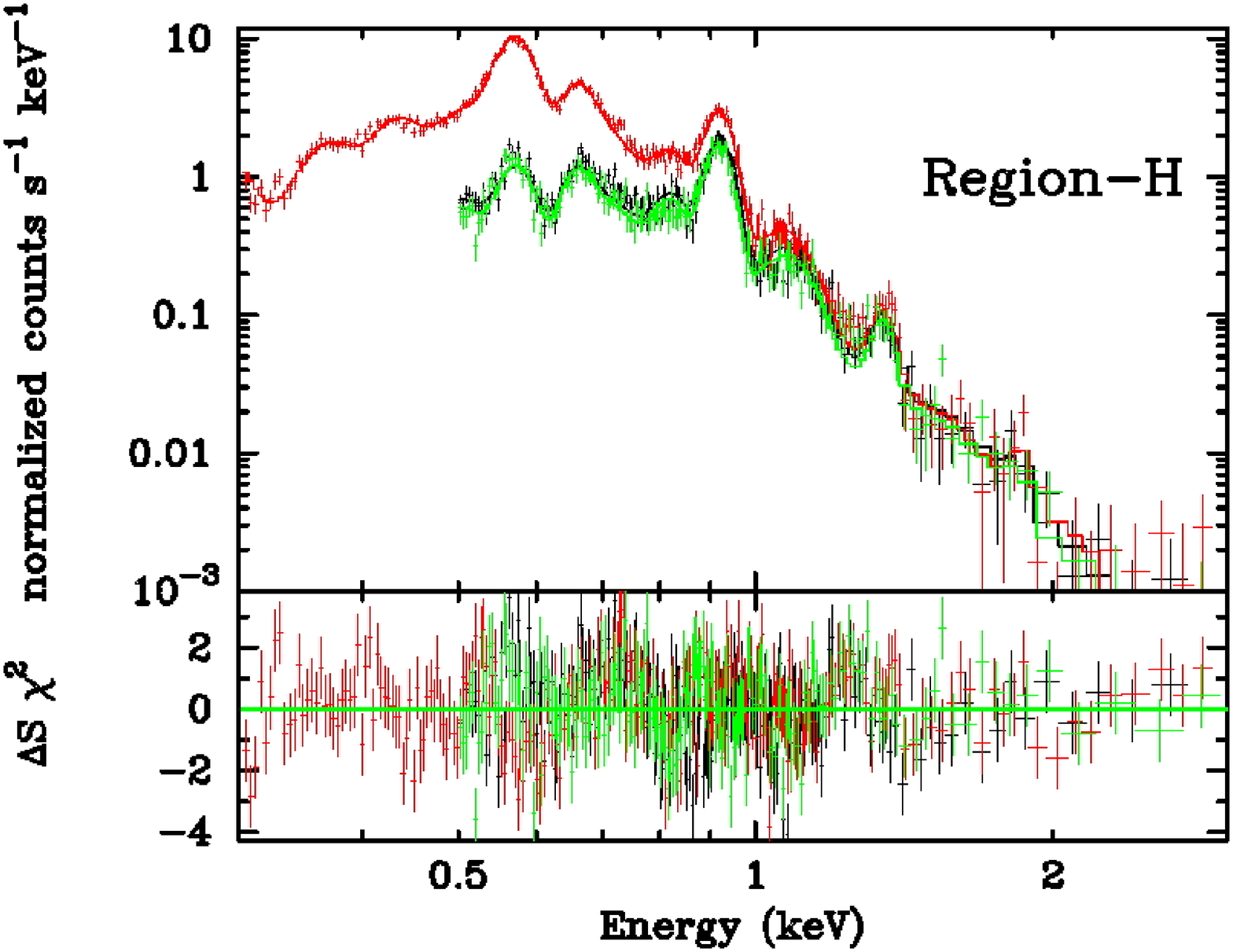}
\includegraphics*[width=8cm, bb=0 0 850 680]{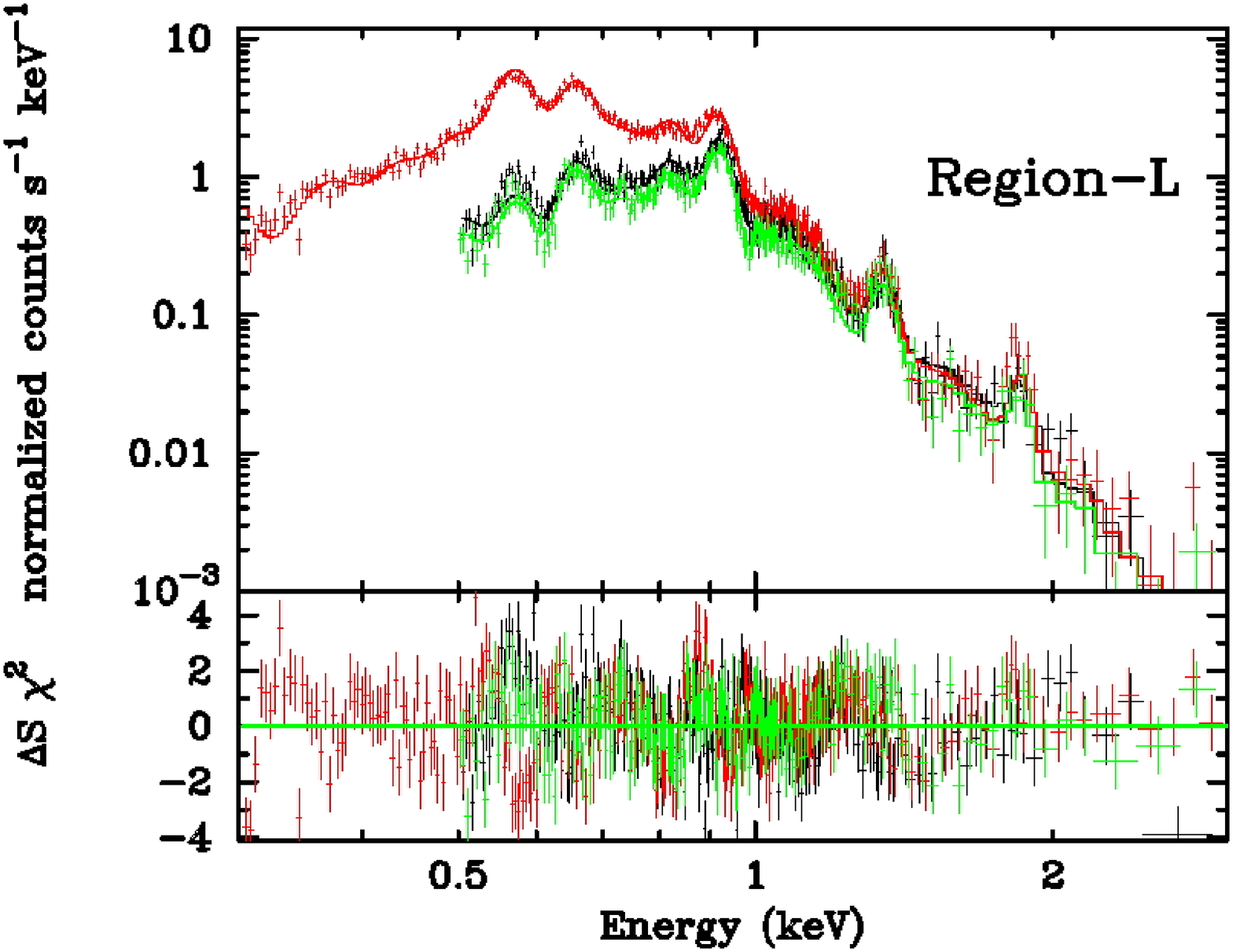}
  \end{center}
  \caption{X-ray spectra for region-H and region-L.}
	\label{fig:spectrum_HL} 
\end{figure}

Fig.~\ref{fig:correlation} shows the variations of various parameters as
a function of angular distance from the shock front.  The value of
$kT_\mathrm{e}$ is almost constant with a slight increase towards
inside.  The value of $\tau$ shows an increase towards inside at the very limb region that can
be seen in the forward shock front (\cite{Miyata1994}).  We noticed that
the metal abundances in the outer annuli are much higher than those in
the inner annuli.  We divide the region into two groups: one is the high
abundance region ({\it region-H}: Ann1 + Ann2 + Ann3) and the other is
the low abundance region ({\it region-L}: Ann7 + Ann8 + Ann9 + Ann10).
The spectra of these two regions are shown in Fig.~\ref{fig:spectrum_HL}
where we can easily see the difference in the line features.  We also
performed the spectral analysis on them by using the VNEI model and
summarized the results in Table~\ref{tab:spectrum_HL}.  
The abundance of O in the region-H is similar to that expected in the
ISM near the Cygnus Loop (O is about 0.4 solar) while that in region-L
is depleted by a factor of two and is similar to the average abundance
in the shell of the Cygnus Loop (O is about 0.2 solar).  The abundances of C, N and Ne are also high/low in the region-H/L while those of Mg and Fe are similar in two regions.  
The width of the
high abundance region is about 3$^\prime$ that is larger than that of
the PSF of the {\it Suzaku} XRT (\cite{Serlemitsos2007}).  The width measured is quite similar to those observed in the NE limb of the Cygnus Loop
(Katsuda et al. 2008a, Katsuda et al. 2008b).

If we assume the spherical symmetry of the Loop, we expect that the
plasma in the outer most region on the apparent image will completely
cover the whole Loop.  Therefore, we expect that the sight line of the
region-L is the superposition of the plasma of the high abundance
region and that of the inside the Loop.  If the low abundance comes
from the projection effect, we can assume that the plasma in the
region-L consists of two components: one is a high abundance thermal
plasma and the other is a very low abundance plasma, resulting an apparent
low abundance.  The low abundance plasma may come from the non-thermal origin.  Then we applied a combination model of VNEI and SRCUT.
The SRCUT model describes a synchrotron radiation from a power law
distribution of electrons with an exponential cut-off at the roll-off
frequency, $\nu_{\rm rolloff}$, (\cite{Reynolds1999}).  If we set the metal abundances to those obtained in the region-H, the reduced $\chi^2$ value increased by 250 (744 d.o.f.).  
If we leave the
abundance free, we find that the depleted abundance without the SRCUT model gives us the best fit result.  In this way, we confirmed that the simple SRCUT model does not explain the low abundance.  This is consistent with the radio result that the SE limb region in the Cygnus Loop shows no radio emission.  A more sophisticated model will be needed.

\section{Discussion}

\subsection{Abundance inhomogeneities in the limb regions of the Cygnus
Loop}

Past X-ray studies have shown that the heavily depleted metal abundance
(O abundance is 0.1--0.2) is commonly observed in the limb of the Cygnus
Loop (\cite{Miyata1994}, \cite{Miyata1998}, \cite{Miyata1999},
\cite{Leahy2004}, \cite{Miyata2007}, \cite{Tsunemi2007},
Katsuda et al.\ 2008a). Among them, Katsuda et al.\ (2008a) reported for the first
time that some area in the very edge of the NE limb show high metal
abundance (C, N, O abundances are $\sim$0.7, $\sim$0.7 and $\sim$0.4).
The high metal abundance in the very edge of the SE limb disclosed here is
quite similar to that of the NE limb, and thus this is another evidence
of abundance inhomogeneity in the limb regions of the Cygnus Loop. This
second discovery may almost deny a possibility that the abundance
inhomogeneity in the NE limb is just an irregular exception, and suggest that
such a feature may be observed in other limb regions under some specific
conditions. 

We here summarize observational facts about the abundance inhomogeneity
regions. The apparent temperature is relatively low (0.3--0.4\,keV) and
almost constant with a slight increase towards inside. The value of $\tau$ clearly shows the non-equilibrium condition of plasma.  The spatial
distribution of $\tau$ also shows an increase from the forward shock to
the inside.  It is quite similar to those obtained in the NE limb regions where the abundances are high (Katsuda et al.\ 2008a).  The value of $\tau$ is large where the abundance is low (inside and bright region) both in the NE limb and in the SE limb.  This is qualitatively understood that the regions with high EM usually show high density and large values of $\tau$.  If we assume the spherical structure and uniform density of the plasma in the SE limb, we can obtain $n_e$ to be 0.4\,cm$^{-3}$.  With taking into account the value of $\tau$, we can expect the elapsed time after the shock heating to be 2500\,years.  This rough estimate surely coincides with the cavity explosion.  Levenson \& Graham (2005) observed the SE region just next to our FOV where its surface brightness is much higher than ours.  They reported the CIE condition at bright region while they also reported the NEI condition at the very limb region showing a similar value of $\tau$ to ours.   

When we see the metal abundance relative to O, the NE limb and the SE
limb look quite similar to each other with an exception of Fe.  In the NE limb, there is a clear difference between the abundance
enhanced region and the abundance depleted region (\cite{Katsuda2008a}).  They can clearly set the boundary between them based on the relative abundances of Fe/O and Mg/O.  Whereas, the boundary between them in the SE limb is not so clear, since we can hardly see a sharp variation of relative abundances in the FOV.  In this way, there are some aspects that are different in two regions.  However, the abundances are quite similar to each other both in the high abundance region and in the low abundance region.  Therefore, we expect that the origin of the plasma in the NE limb is quite similar to that of the SE limb.  

The low metal abundances may be due to the combination of a thin thermal plasma with normal abundance and a non-thermal plasma.  However, our data clearly supported the low abundance plasma rather than a contribution of a simple non-thermal plasma.    

\subsection{Comparison with the radio morphology}

The {\it Suzaku} observation of the NE limb consists of four pointings (NE1 to NE4 from the south to the north) (Katsuda et al., 2008a).  The high abundance region is the very limb in NE3 and NE4.   Their results indicate that the abundance in the NE4 limb is higher than that in the NE3 limb.  According to the radio map at 1420\,MHz (\cite{Uyaniker2004}), the brightness temperature, T$_B$, at NE1 and NE2 is one of the highest (2--10\,K) in the Cygnus Loop while that in NE3 is around 1\,K and that in NE4 is less than 0.3\,K.  Therefore, we find that the region where T$_B$ is low can show high abundance in X-ray spectrum.

Any line of sight through the remnant almost certainly intersects
various emitting regions.  However, the line of sight near the limb can intersect the very limited plasma condition.  Therefore, we speculate that the very limb region in NE3 and NE4 contains only the thin hot plasma with high abundance where the radio intensity is weak.  We can expect that the limb region with weak radio intensity shows high abundance that is much higher metal abundance than that in its surrounding region.   The radio intensity in the Cygnus Loop is quite strong with some exceptions.  One exception is the NE limb (NE3 and NE4 and its north) and the other is the SE limb.  In particular, no radio intensity is reported in the SE limb (\cite{Uyaniker2004}).  We found that both limbs contain regions showing relatively high abundance (O is about 0.4 solar) while the inner regions show depleted abundance (O is about 0.2).  The NE limb and the SE limb are similar in abundance structure while they are quite different in radio structure. 

\section{Conclusion}
 
We observed the SE limb of the Cygnus Loop with {\it Suzaku} where the surface brightness is relatively weak and smooth.  We found that the limb region (region-H) shows relatively high abundance (O$\sim 0.4$) while that in the inner region (region-L) shows low abundance (O$\sim 0.2$).  The abundance of O in region-H is similar to that of the ISM near the Cygnus Loop while that in region-L is similar to that in the Cygnus Loop observed so far.  The high metal abundance in the SE limb is quite similar to that of the NE limb, and thus this is another evidence of abundance inhomogeneity in the limb regions of the Cygnus Loop.  

The radio map shows a quite different feature: the NE limb is in the radio bright region while the SE limb shows almost no radio.  Based on the radio intensity morphology on the NE limb, we noticed that the high abundance in the very limb region (NE3 and NE4) corresponds to the weak radio intensity while the low
abundance in the very limb region (NE1 and NE2) corresponds to the high radio intensity.  However, there is no radio intensity reported in the SE limb we observed.  Therefore, the non-thermal emission does not contribute to the low metal abundance both in the NE limb and in the SE limb that are left as an open question.

\section*{Acknowledgements}

This work is partly supported by a Grant-in-Aid for Scientific
Research by the Ministry of Education, Culture, Sports, Science and
Technology (16002004).  This study is also carried out as part of the
21st Century COE Program, \lq{\it Towards a new basic science: depth
  and synthesis}\rq.  H.U. and S.K. are supported by JSPS Research
Fellowship for Young Scientists. 



\newpage

\begin{table*}
  \begin{center}
  \caption{Spectral-fit parameters for annular regions}\label{tab:ex_param}
   \begin{tiny}
    \begin{tabular}{lcccccccccc}
      \hline\hline
Parameter & Ann1 & Ann2 & Ann3 & Ann4 & Ann5 & Ann6 & Ann7 & Ann8 & Ann9 & Ann10\\
\hline
$kT_\mathrm{e}$[keV] \dotfill 
	& 0.35$\pm 0.01$ & 0.34$\pm 0.01$ & $0.38 \pm 0.01$
	&$0.34 \pm 0.01$& $0.34\pm 0.01$ & $0.35 \pm 0.01$
	&$0.35 \pm 0.01$& $0.38\pm 0.01$ & $0.42 \pm 0.01$& $0.41 \pm 0.01$\\
	
C(=Si=S) \dotfill
	& $0.60\pm 0.05$ & $0.93\pm0.06$ & $0.89\pm0.05$ & $0.56\pm0.04$ & $0.47\pm0.04$
	& $0.34\pm 0.03$ & $0.26\pm 0.03$ & $0.39\pm0.04$ & $0.40 \pm0.05$ & $0.26\pm 0.03$ \\
N \dotfill
	& $0.85\pm0.06$ & $1.08\pm 0.06$ & $0.78\pm0.05$ & $0.52\pm0.03$ & $0.32\pm0.03$ 
	& $0.24\pm0.02$ & $0.24 \pm 0.02$ & $0.27\pm0.03$ & $0.40\pm 0.04$ & $0.24\pm0.03$ \\
O \dotfill
	& $0.44\pm0.02$ & $0.55\pm0.02$ & $0.43\pm 0.02$ & $0.31\pm 0.01$ & $0.21\pm 0.01 $ & $0.18\pm 0.01$
	& $0.16\pm 0.01$ & $0.19\pm 0.01$ & $0.22\pm 0.01$ & $0.17\pm 0.01$ \\
Ne \dotfill
	& $0.69\pm 0.03$ & $0.92\pm 0.03$ & $0.70\pm 0.02$ & $0.54\pm 0.02$ & $0.38\pm 0.01$ 
	& $0.34\pm 0.01$ & $0.30\pm 0.01$ & $0.34\pm 0.01$ & $0.46\pm 0.02$ & $0.33 \pm 0.01$\\
Mg \dotfill
	& $0.5\pm 0.1$ & $0.47\pm 0.09$ & $0.22\pm 0.04$ & $0.29\pm 0.04$ 
	& $0.18\pm 0.03$ & $0.20\pm 0.03$ & $0.15\pm 0.03$ 
	& $0.16^\pm 0.03$ & $0.21\pm 0.03$ & $0.18\pm 0.03$ \\
Fe(=Ni) \dotfill
	& $0.52\pm 0.04$ & $0.39\pm 0.02$ & $0.27\pm 0.01$ & $0.26\pm 0.01$
	& $0.19\pm 0.01$ & $0.18\pm 0.01$ & $0.17\pm 0.01$ & $0.19\pm 0.01$
	& $0.23\pm 0.01 $ & $0.19\pm 0.01$ \\
log$(\tau /\mathrm{cm}^{-3}\,\mathrm{sec})$\dotfill
     & $10.25\pm 0.02$ & $10.39\pm 0.02$ & $10.52\pm 0.02$ & $10.54\pm 0.02$ 
     & $10.65\pm 0.02$ & $10.64\pm 0.02$ & $10.7\pm 0.02$ & $10.6\pm 0.02$ 
     & $10.6\pm 0.02$ & $10.72\pm0.02$ \\
EM[$\times10^{18}$ cm$^{-5}$]\dotfill
	& $3.92\pm 0.05$ & $4.14\pm 0.04$ & $5.41\pm 0.05$ & $6.81\pm 0.06$ 
	 & $10.1\pm 0.1$ & $11.1\pm 0.1$ & $12.4\pm 0.1$ & $9.3\pm 0.1$ & $7.0\pm 0.1$ & $11.5\pm 0.1$ \\
\hline
$\chi^2$/d.o.f. \dotfill
 & 392/310 & 316/254 & 429/302 & 457/313 & 390/302 & 425/298 & 433/284 & 444/257 & 343/232 & 578/444 \\
      \hline
\\[-8pt]
  \multicolumn{3}{@{}l@{}}{\hbox to 0pt{\parbox{140mm}{\footnotesize
     \par\noindent 
\footnotemark[$*$]Si/S are fixed to O and other elements are fixed to those of solar values.\\
     The values of abundances are multiples of solar value.\\  The errors are in the range $\Delta\,\chi^2\,<\,2.7$ on one parameter.  \\ 
\par\noindent 
}\hss}}
    \end{tabular}
   \end{tiny}
  \end{center}
\end{table*}

\begin{table*}
  \begin{center}
  \caption{Spectral-fit parameters for high and low abundance regions}\label{tab:spectrum_HL}
    \begin{tabular}{lccc}
      \hline\hline
Parameter & high region & low region \\
model    &  VNEI  &  VNEI    \\
\hline
$kT_\mathrm{e}$[keV] \dotfill 
	& 0.34$\pm 0.01$ & 0.39$\pm0.01$ \\
C \dotfill
	&0.70$\pm 0.03$ & 0.38$\pm 0.02$  \\
N \dotfill
	& 0.81$\pm 0.03$ & 0.33$\pm 0.02$ \\
O \dotfill
	& 0.44$\pm 0.01$& $0.19 \pm 0.01$  \\
Ne \dotfill
	& 0.72$\pm 0.01$ & 0.38$\pm0.01$ \\
Mg \dotfill
	& 0.28$\pm 0.04$& $0.24\pm0.02$  \\
Fe(=Ni) \dotfill
	&0.30$\pm 0.01$ & $0.25\pm0.01$ \\
log$(\tau /\mathrm{cm}^{-3}\,\mathrm{sec})$\dotfill
     &10.42$\pm 0.01$ & 10.77$\pm 0.02$  \\
EM[$\times10^{19}$ cm$^{-5}$]\dotfill
	& $0.43 \pm0.01 $& $1.03 \pm0.01$  \\
\hline
$\chi^2$/d.o.f. \dotfill 
	& 1008/650 & 1208/733 \\
      \hline
\\[-8pt]
  \multicolumn{3}{@{}l@{}}{\hbox to 0pt{\parbox{140mm}{\footnotesize
     \par\noindent 
\footnotemark[$*$]Si/S are fixed to O and other elements are fixed to those of solar values.\\
     The values of abundances are multiples of solar value.\\  The errors are in the range $\Delta\,\chi^2\,<\,2.7$ on one parameter.  \\ 
\par\noindent 
}\hss}}

    \end{tabular}
  \end{center}
\end{table*}

\end{document}